\documentclass[aps,prl,showpacs,reprint,amsmath,amssymb,floatfix,groupedaddress]{revtex4-1}

\usepackage{graphicx}
\usepackage{epsfig}

\begin{document}

\title{Enhancement of Ultracold Molecule Formation Using Shaped Nanosecond Frequency Chirps}

\author{J. L. Carini$^{1}$}
\author{S. Kallush$^{2,3}$}
\author{R. Kosloff$^{3}$}
\author{P. L. Gould$^{1}$}
\affiliation{$^{1}$Department of Physics, University of Connecticut, Storrs, Connecticut 06269, USA \\ $^{2}$Department of Physics and Optical Engineering, ORT Braude, P.O. Box 78, Karmiel, Israel \\ $^{3}$Department of Physical Chemistry and the Fritz Haber Research Center for Molecular Dynamics, The Hebrew University, 91094, Jerusalem, Israel}

\date{\today}

\begin{abstract}
We demonstrate that judicious shaping of a nanosecond-time-scale frequency chirp can dramatically enhance the formation rate of ultracold $^{87}$Rb$_{2}$ molecules. Starting with ultracold $^{87}$Rb atoms, we apply pulses of frequency-chirped light to first photoassociate the atoms into excited molecules and then, later in the chirp, de-excite these molecules into a high vibrational level of the lowest triplet state, \textit{a} $^{3}\Sigma_{u}^{+}$. The enhancing chirp shape passes through the absorption and stimulated emission transitions relatively slowly, thus increasing their adiabaticity, but jumps quickly between them to minimize the effects of spontaneous emission. Comparisons with quantum simulations for various chirp shapes support this enhancement mechanism.
\end{abstract}


\maketitle


The fields of coherent control and ultracold physics are often considered orthogonal. Coherent control \cite{Rice92,Brumer92,Brif10} usually deals with the manipulation of internal degrees of freedom, such as electronic state, or in the case of molecules, vibration and rotation. This manipulation is usually done on ultrafast time scales in order to keep pace with the internal dynamics. In contrast, with ultracold atoms \cite{Metcalf99} and molecules \cite{Krems09,Jin12,Carr15}, the focus is usually on the external degrees of freedom, such as position and momentum, and the corresponding time scales are much slower because of the low temperatures. A phenomenon which provides some overlap between these two areas is ultracold photoassociation \cite{Jones06}, whereby two free atoms absorb one or more photons and are bound into a molecule. In this elementary chemical reaction, catalyzed by the photon(s), the external degrees of freedom of the colliding atoms are replaced, in the center-of-mass frame, by molecular vibration and rotation. Because the initial continuum energy at ultracold temperatures is well defined, coherent interactions can be important in this bond formation process. 

	Photoassociative formation of ultracold molecules is usually done with continuous light, relying on spontaneous emission (SPE) to convert the electronically-excited molecules to a lower-lying state, in our case the \textit{a} $^{3}\Sigma_{u}^{+}$ metastable triplet state of Rb$_{2}$. This SPE is not only incoherent, but also distributes the population into a number of vibrational states and back into the continuum. With a view towards various applications \cite{Krems09} in ultracold chemistry, tests of fundamental physics, and quantum information, there have been a number of proposals \cite{Koch12,Vala00,Luc-Koenig04a,Luc-Koenig04b,Koch06a,Poschinger06,Koch06b,Brown06a,Koch06c,Mur-Petit07,Kallush07a,Kallush08,Koch08,Koch09,Kallush07b,Huang14,Tomza12} to apply ultrafast coherent control techniques to ultracold molecule formation in order to coherently form the molecules in a designated target state. The success in this endeavor has, so far, been limited. The opposite process, photodestruction of already existing ultracold molecules, has been optimized with ultrafast coherent control \cite{Salzmann06,Brown06}, and coherent transients have been observed in ultrafast photoassociation \cite{Salzmann08,McCabe09}.
	
On much slower time scales, we recently used nanosecond pulses of frequency-chirped light for photoassociation \cite{Carini13}. The main finding was a significant dependence on chirp direction, attributed to coherent stimulated emission (STE) following the initial photoassociation. However, the molecular formation rates with linearly chirped light were still less than those with unchirped pulses of the same duration and pulse energy. In the present experiment, we use faster and broader chirps and higher intensities. Most importantly, inspired by our recent calculations using local control \cite{Carini15}, we incorporate shaping of the chirp as a means of increasing the molecule formation rate. Taken together, these improvements enhance the contribution of coherent STE while diminishing the role of incoherent SPE in the molecule formation process. This yields not only an increased contrast between positive and negative chirps, but also an improved performance relative to unchirped pulses.

	Similar to reference \cite{Carini13}, the experiment starts with ultracold (150 $\mu$K) and dense (8x10$^{10}$ cm$^{-3}$) $^{87}$Rb atoms which are then illuminated with frequency-chirped light in order to photoassociate (PA) them into ultracold $^{87}$Rb$_{2}$ molecules, as shown in Fig. 1. The PA light is centered on a transition to 0$_{g}^{-}$ (v'=78), 7.79 GHz below the 5S$_{1/2}$(F=2) $\rightarrow$ 5P$_{3/2}$(F'=3) atomic transition \cite{Kemmann04,Carini13}. Some fraction of these excited molecules undergo either SPE or STE into high vibrational levels of the \textit{a} $^{3}\Sigma_{u}^{+}$ metastable state and are detected by resonance-enhanced multiphoton ionization (REMPI). Of particular interest is v''=39, the next-to-last vibrational level of \textit{a} $^{3}\Sigma_{u}^{+}$, which is sufficiently weakly bound that it can be populated by STE from the same chirp driving the PA step.	
	
		In order to produce fast and shaped frequency chirps, we use a high-speed  electro-optical phase modulator, situated in a fiber loop and  driven by an arbitrary waveform generator \cite{Rogers07}. A cw external-cavity diode laser, with $\sim$1 MHz linewidth, provides the initial seed light. The resulting frequency-chirped light injection locks a high power diode laser whose output is controlled with an acousto-optical modulator to yield a Gaussian intensity pulse with a measured FWHM of 15 ns. We note that our time-domain chirp generation is quite different from ultrafast frequency-domain shaping techniques \cite{Weiner00}. In the former, we maintain the temporal duration and peak intensity of the pulse while increasing its bandwidth. In the latter, the bandwidth is not increased, but the pulse is stretched in time and its peak intensity reduced. 
		
\begin{figure}
    \centering 
    \includegraphics[width=8.6cm]{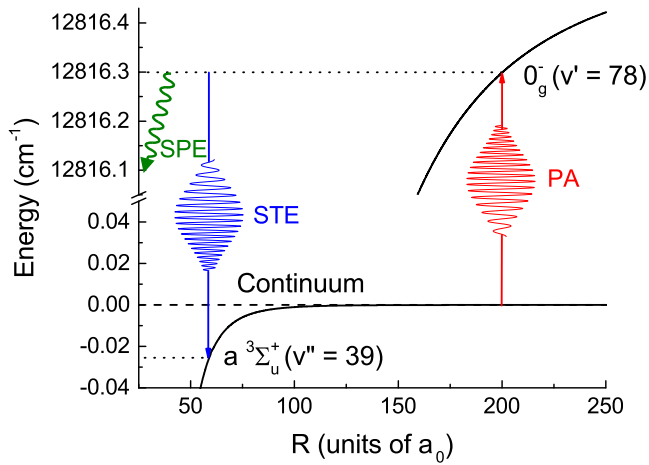} 
    \caption{(Color online) Long-range portion of molecular potentials and the vibrational levels relevant to the frequency-chirped molecule formation. Transitions between the lower \textit{a} $^{3}\Sigma_{u}^{+}$ state and 0$_{g}^{-}$ excited state are also shown: photoassociation (PA) and stimulated emission (STE) induced by the chirped pulse; and spontaneous emission (SPE) of the excited state.}
\end{figure}
	
\begin{figure}
    \centering 
    \includegraphics[width=8.0cm]{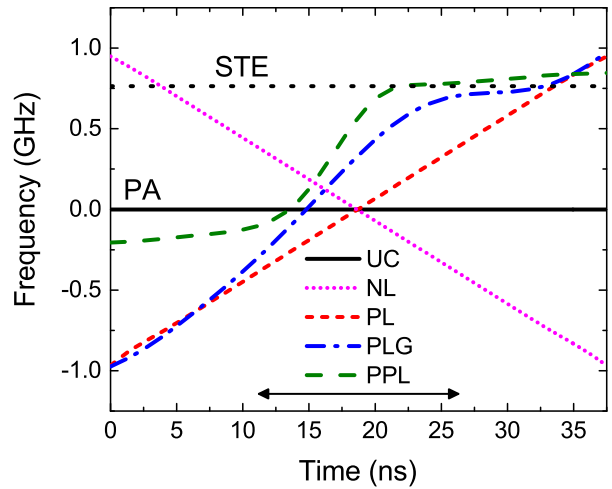} 
    \caption{(Color online) Measured frequencies vs. time, smoothed with a 2 ns FWHM Gaussian, for the various chirps used for molecule formation: unchirped (UC); negative linear (NL); positive linear (PL); positive linear plus Gaussian (PLG); and positive piecewise linear (PPL). The solid and dotted horizontal lines represent the PA transition to 0$_{g}^{-}$ (v'=78) and the STE transition to \textit{a} $^{3}\Sigma_{u}^{+}$ (v''=39), respectively, as shown in Fig. 1.  The timing of the intensity pulse is indicated by the double-ended arrow (length = 15 ns FWHM).}
\end{figure}

The various chirp shapes used in the present work, measured with a heterodyne technique, are shown in Fig. 2: linear chirps, both positive (PL) and negative (NL), with slopes of $\sim$1.9 GHz in 37.5 ns; a positive linear chirp with a Gaussian (0.425 GHz amplitude, 15 ns FWHM) superimposed (PLG); and a positive piecewise linear (PPL) chirp comprising gently sloping ($\sim$10 MHz/ns) initial and final segments with a steep ($\sim$120 MHz/ns) rise in between. The central temporal region of each chirp is selected with the 15 ns FWHM Gaussian intensity pulse. For the linear chirps, these chirp times and pulse widths are a factor of $\sim$2.7 shorter than in our earlier work, a key factor in tipping the balance between incoherent and coherent processes. Also the chirp range is a factor of $\sim$2 wider, allowing the coherent combination of absorption and STE to occur within the high intensity portion of the pulse. 

	The REMPI detection employs 5 ns, $\sim$4.8 mJ pulses from a pulsed dye laser operating at 10 Hz. The Gaussian atomic cloud has an average 1/e$^{2}$ radius of 172 $\mu$m, while the REMPI beam is larger, $\sim$3 mm in diameter, in order to more effectively overlap the ballistically expanding ultracold molecules. The wavelength is centered on a broad feature in the spectrum at $\sim$16608.5 cm$^{-1}$ which, based on previous work \cite{Kemmann04,Gabbanini00}, ionizes high vibrational levels of the \textit{a} $^{3}\Sigma_{u}^{+}$ state. Due to the $\sim$0.2 cm$^{-1}$ laser bandwidth, these high-v'' levels are not resolved. Ions are accelerated into a Channeltron detector and measured with a digital boxcar averager which distinguishes Rb$_{2}^{+}$ from background Rb$^{+}$ by time of flight.
	
	The chirped pulses, repeated at 2.2 MHz, not only form ultracold \textit{a} $^{3}\Sigma_{u}^{+}$ molecules at a time-averaged rate R, but can also photodestroy already existing molecules at a time-averaged rate per molecule $\Gamma$$_{PD}$. In addition, the molecules escape ballistically from the detection region at a rate $\Gamma$$_{esc}$. The steady-state number of detectable molecules is a balance between the formation and total loss: N$_{SS}$ = R/($\Gamma$$_{PD}$ + $\Gamma$$_{esc}$). To extract R, the quantity of interest, we determine the total loss rate, $\Gamma$$_{PD}$ + $\Gamma$$_{esc}$, by measuring the exponential approach to steady-state as we vary the formation time (number of chirped pulses) prior to the REMPI pulse. We independently measure $\Gamma$$_{esc}$ = 100(4) s$^{-1}$ using molecules produced by MOT light \cite{Carini13}. 

	We plot the molecular formation rate R as a function of peak intensity for the various chirps in Fig. 3. Several trends are immediately obvious. First, as noted in our previous work on slower time scales, the PL chirp significantly outperforms the corresponding NL chirp. Second, the PL chirp does as well as, or perhaps even outperforms, the unchirped pulse. This is in contrast to our earlier work, where the unchirped pulse significantly outperformed both the PL and NL chirps. Third, the PLG chirp does about as well as the PL chirp. Finally, the PPL chirp, inspired by our local control simulations \cite{Carini15}, dramatically outperforms all other chirp shapes. This demonstrates the advantage of tailoring the shape of the chirp to match the system dynamics.

\begin{figure}
    \centering 
    \includegraphics[width=8.0cm]{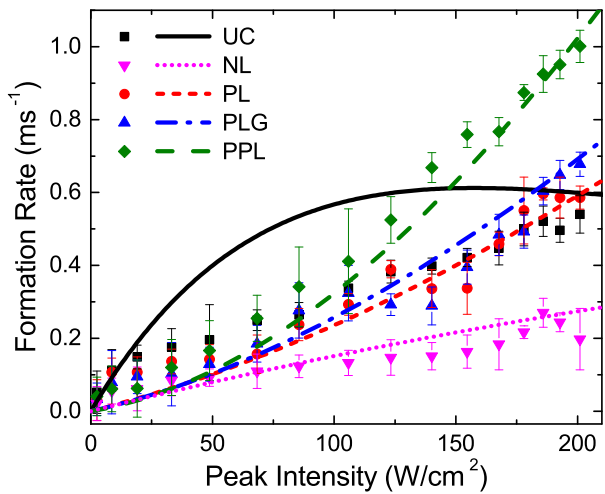} 
    \caption{(Color online) Measured (points) and simulated (curves) molecule formation rates vs. peak intensity for the various chirps in Fig. 2. The theory curves have been scaled by a factor of 1.997.}
\end{figure}

	In order to understand the details of the frequency-chirped molecular formation, we perform quantum simulations of the ultracold collisional dynamics. These are described more completely in earlier publications \cite{Carini13,Carini15}; here we provide a brief summary. We solve the time-dependent Schr$\ddot{o}$dinger equation with a dressed-state Hamiltonian which incorporates the \textit{a} $^{3}\Sigma_{u}^{+}$ metastable state (including its low-energy continuum) and the 0$_{g}^{-}$ and 1$_{g}$ long-range excited states, as assigned in earlier work \cite{Kemmann04,Carini13,Carini15}. The laser-induced coupling between these states is time dependent due to variations of both the frequency (chirp) and the amplitude (pulse envelope). The calculations are done in a restricted basis set of vibrational levels and continuum states determined by the mapped Fourier grid method \cite{Kallush06}. Although SPE is less important on the faster time scales of the present work, we still incorporate it using multiple sink channels, including to the continuum. 	The lifetimes of 0$_{g}^{-}$ and 1$_{g}$ are taken to be 26.2 ns and 22.8 ns, respectively \cite{Julienne91,Gutterres02}.  Assuming an initial box-normalized scattering state, the calculation tracks the normalized population in each state. These state probabilities are converted to the time-averaged formation rate at a fixed peak intensity by: 1) accounting for the thermal ensemble, as described in \cite{Koch06b}; 2) multiplying by the chirp repetition rate; 3) averaging over the Gaussian atomic density distribution; 4) averaging over the Gaussian laser profile (average 1/e$^{2}$ radius of 130 $\mu$m); and 5) summing over partial waves up to J=5. As in our earlier work \cite{Carini13,Carini15}, we exclude the contributions of the barely-bound (39 MHz) and easily photodestroyed \textit{a} $^{3}\Sigma_{u}^{+}$ (v''=40) level.
	
	The resulting formation rates, plotted as functions of the peak intensity, are shown together with the data in Fig. 3. These rates are based on populations 200 ns after the start of the chirp, in order to allow excited states to decay. We have scaled the theory curves by a factor of 1.997 which provides the best match of the chirped curves to the data.  Such a scaling is not unreasonable given the factor of $\sim$2 systematic uncertainty in the absolute atomic density. Except for the unchirped pulses, the curves for the different chirp shapes match the experimental data quite nicely. In particular, the NL chirp gives the lowest rate, the PPL the highest rate, and the PL and PLG give similar rates and are sandwiched in between. 

	The key point of the present work is the dramatically enhanced formation rate realized with an appropriately shaped frequency chirp, the PPL curve in Fig. 3. This chirp shape was inspired by our recent simulations using local control of the phase to optimize the molecular formation \cite{Carini15}. Local control \cite{Marquetand07,Engel09} is a specific implementation of coherent control which relies on a unidirectional and noniterative time propagation scheme. The field is adjusted at each time step in order to optimize the target, in our case molecule formation, at the next step. The optimal temporal variation of frequency which emerged from these simulations was a rapid (subnanosecond) jump between two frequencies, the first being the PA transition up to 0$_{g}^{-}$ (v'=78), and the second being the STE transition from 0$_{g}^{-}$ (v'=78) down to \textit{a} $^{3}\Sigma_{u}^{+}$ (v''=39). This step-function chirp significantly outperformed two simultaneous unchirped frequencies, resonant with the PA and STE transitions, for the same total intensity. Due to speed limitations of our chirp production technique, uncertainties in the exact transition frequencies, and possible slow drifts of the center frequency of the chirp, we have utilized the positive piecewise linear (PPL) chirp in place of the step function. As seen in Fig. 2, this PPL chirp initially ramps slowly through the first (PA) transition, then increases its slope in order to arrive quickly near the second (STE) transition, and finally ramps slowly through this second transition. The slow ramps through each transition are more adiabatic, resulting in higher efficiency, while the reduced time spent between them minimizes the probability of SPE. This reduced time also makes both transitions resonant at times closer to the peak intensity of the pulse.   
		
The difference between positive and negative linear chirps has been discussed previously \cite{Carini13}, and because of our higher intensities and shorter time scale, is even more pronounced here. The higher formation rate for the positive chirp is due to the optimal time ordering of the two transitions. The PA transition is resonant first, followed by the STE transition. For the negative chirp, this order is reversed, and consequently there is very little excited-state population available to be driven down to \textit{a} $^{3}\Sigma_{u}^{+}$ (v''=39) by STE. 

Another obvious improvement over our previous measurements is that the formation rates for the positive chirps match or exceed that for the unchirped pulse. In the simulations, the rates for the positive chirps grow quadratically with intensity, consistent with a two-photon process (PA followed by STE). The rate for the negative chirp is rather linear in intensity, as would be expected for a one-photon process (PA followed by SPE). Adding the Gaussian to the positive linear chirp does not result in much improvement in the formation rate. 

\begin{figure}
    \centering 
    \includegraphics[width=8.0cm]{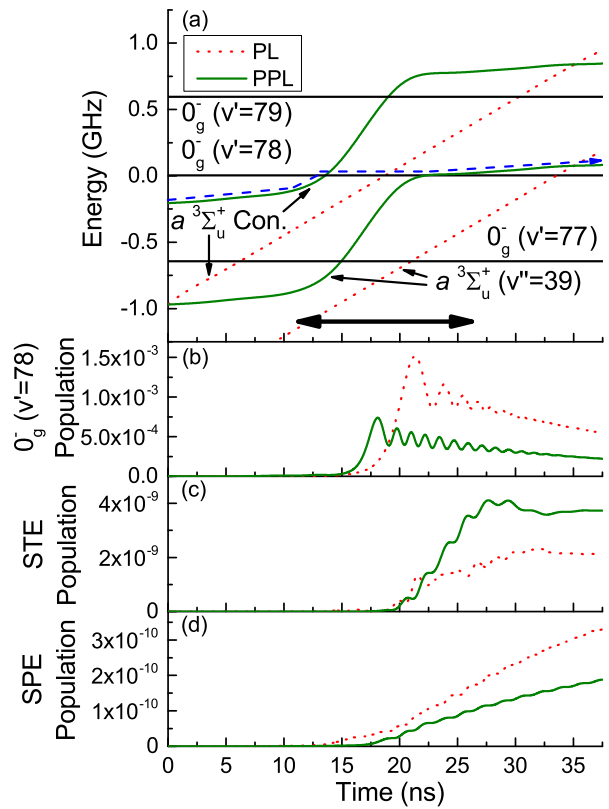} 
    \caption{(Color online) (a) Temporal evolution of dressed-state molecular energies for the cases of the positive linear (PL) and positive piecewise linear (PPL) chirps. Horizontal lines indicate the energies of the v'=77-79 vibrational levels of the 0$_{g}^{-}$ excited state. Upper and lower sloped lines indicate the energies of the continuum and the v''=39 bound level of the \textit{a} $^{3}\Sigma_{u}^{+}$ state, respectively, with the energy of the chirped photon added. The dashed line denotes the adiabatic path from the free-atom continuum to the bound v''=39 molecule for the PPL chirp. The double-ended arrow (length = 15 ns FWHM) indicates the timing of the intensity pulse. Time-dependent populations for a peak pulse intensity of 150 W/cm$^{2}$: (b) v'=78 excited state; (c) v''=39 populations resulting from stimulated emission (STE); (d) v''=39 populations resulting from spontaneous emission (SPE). At long times, the excited states have completely decayed and the SPE populations in (c) reach 6x10$^{-10}$ and 3x10$^{-10}$ for the PL and PPL chirps, respectively. For the PPL chirp, note the enhanced population from STE and diminished population from SPE relative to the PL chirp.}
\end{figure}

To gain some insight into the relative performance of the various chirps, we examine how the state populations evolve in time. An example is shown in Fig. 4, where we compare the PL and PPL chirps. In Fig. 4(a), the energy levels are shown in the dressed picture, where the time-dependent photon energy is added to the continuum and v''=39 level of the \textit{a} $^{3}\Sigma_{u}^{+}$ lower state. Note that all \textit{a} $^{3}\Sigma_{u}^{+}$, 0$_{g}^{-}$, and 1$_{g}$ levels in our restricted basis set are included in the calculation, but for clarity, only a few selected levels are shown. A crossing between a dressed lower level and an excited level (horizontal line) occurs when the chirp is resonant with a transition between these levels. The small avoided crossings due to the coupling by the laser field are not shown. If traversal through these avoided crossings were completely adiabatic, which is not the case at our intensities, the dashed curve would be followed for the PPL chirp.

The lower panels of Fig. 4 show the time-dependent populations for a fixed intensity and collision energy, and a single partial wave (J=0). For both chirps, the 0$_{g}^{-}$ (v'=78) excited-state population appears near the first curve crossing (PA resonance) and the transfer to \textit{a} $^{3}\Sigma_{u}^{+}$ (v''=39) by STE occurs somewhat later. Unlike in our previous work on slower time scales, the STE contribution overwhelms the SPE contribution, especially for the PPL chirp where it is larger by at least an order of magnitude. Interestingly, the final \textit{a} $^{3}\Sigma_{u}^{+}$ (v''=39) population for the PPL chirp greatly exceeds that for the PL chirp, despite its excited-state population being significantly smaller. This emphasizes the advantages of the shaped PPL chirp: the steep portion reaches the STE resonance sooner, and therefore at higher intensity and with less loss from SPE, while the gentle slope near the resonance yields a more adiabatic, and therefore more efficient, downward transition.

It is interesting to compare our results with the recent work on ultrafast coherent control of bond formation at higher temperatures \cite{Levin15}. Although the overall goals were similar, there are major differences between the experiments. Their timescales and temperatures differ by about five and seven orders of magnitude, respectively, from those in our work. Their target state was an excited one, whereas ours is the lowest (metastable) triplet state. Although a significant enhancement was seen for positive vs. negative chirps in both cases, their mechanism was much more complicated. It involved a combination of Franck-Condon filtering, chirp-dependent Raman transitions, and coherent vibrational dynamics. The contrasting simplicity of our mechanism results from our much lower temperature.  Despite these differences, it is satisfying that coherent bond formation has been realized in these extremely disparate regimes.

	In summary, we have utilized a shaped frequency chirp to enhance the photoassociative production of ultracold molecules. The following sequence: slow chirp through the photoassociation resonance, followed by a rapid jump to a resonance connecting to the target state, and finally a slow passage through this second resonance, provides a significant improvement over a positive linear chirp. Results of quantum simulations show good agreement with the measurements. Further improvements are expected for faster chirps, shorter pulses, higher intensities, and empirically optimized chirp shapes. Because of the quadratic dependence, increasing the intensity will be particularly beneficial.

\begin{acknowledgments}
	This work is supported by the U.S. Department of Energy Office of Science, Office of Basic Energy Sciences, Chemical Sciences, Geosciences, and Biosciences Division under Award Number DE-FG02-92ER14263. We also acknowledge support from the US-Israel Binational Science Foundation through grant number 2012021. 
\end{acknowledgments}

\bibliography{library}
\end{document}